\RequirePackage[2020-02-02]{latexrelease}
\documentclass[nofootinbib]{revtex4}

\usepackage{amsmath}
\usepackage{amssymb}

\begin{document}
\title{Forms of BRST Symmetry on a Prototypical First-class System}
\author{Sumit Kumar Rai \footnote{e-mail address:
\ \ sumitssc@gmail.com}}
\affiliation{ Department of Physics,\\
Sardar VallabhBhai Patel College,\\
(A constituent unit of V.K.S.U, Ara),Kaimur-Bhabua,821101 INDIA. \\
}
\author{Bhabani Prasad Mandal\footnote{e-mail address: \ \ bhabani@bhu.ac.in, \ \ bhabani.mandal@gmail.com  } }
\affiliation{ Department of Physics,\\
Banaras Hindu University,\\
Varanasi-221005, INDIA. \\
}
\author{Ronaldo Thibes \footnote{e-mail address: \ \ thibes@uesb.edu.br}}
\affiliation{Departamento de Ci\^encias Exatas e Naturais,\\
Universidade Estadual do Sudoeste da Bahia,\\
Itapetinga - 45700000, Brazil.\\
}

\begin{abstract}
We obtain the various forms of BRST symmetry by using the Batalin-Fradkin-Vilkovisky formalism in a prototypical first class system. We have shown that the various forms of symmetry can be obtained through canonical transformation in the ghost sector.  The so called "dual-BRST" symmetry which is claimed to be an independent symmetry due to its roots in differential geometry is obtained from usual BRST symmetry by making a canonical transformation in the ghost sector. 
\end{abstract}
\maketitle
\section{Introduction}
The Becchi-Rouet-Stora and Tyutin (BRST) symmetry \cite{brst} forms the foundation of modern gauge theory quantization and serves as a crucial tool in characterizing various renormalizable field-theoretic models. Path integral quantization of gauge theories \cite{hete} can be carried out using both the Lagrangian and Hamiltonian formulations. In both approaches, the phase space is extended by incorporating Grassmannian odd ghost variables, ensuring gauge invariance through BRST symmetry.

The Hamiltonian formalism, developed by Batalin, Fradkin, and Vilkovisky (BFV) \cite{frvi}, provides a powerful method for studying BRST quantization in constrained systems. 
The BFV approach offers several advantages: it does not require the off-shell closure of the gauge algebra and thus eliminates the need for auxiliary fields; it relies heavily on BRST transformations that remain independent of the gauge condition; and it applies to Lagrangians that are not necessarily quadratic in velocities, making it more general than the strict Lagrangian approach. Being based on the Hamiltonian formulation, it aligns more closely with Hilbert space techniques and unitarity. This method employs an extended phase space where the Lagrange multipliers and ghost variables are treated as dynamical entities. The generator of BRST symmetry for systems with first-class constraints can be constructed in a gauge-independent manner, with its cohomology defining the physical states.
Extensive research has been conducted using the BFV approach for systems with first-class constraints, such as QED and U(1) gauge theory \cite{gara,nega,gaete}.  Additionally, different forms of BRST symmetry\cite{mara,upma,pama,pama1,mara1,pama2} have been explored, including non-local and non-covariant symmetry in QED \cite{lamc}, non-local but covariant symmetry in QED \cite{tafi}, and local but non-covariant symmetry in Abelian gauge theories \cite{yale}. In all these cases, the variation of the gauge-fixing term vanishes, which is considered "dual" to the vanishing of the variation of the kinetic term in conventional BRST symmetry. The so-called dual-BRST (or co-BRST) symmetry, in which the gauge-fixing term remains independently invariant while the kinetic term variation cancels with that of the ghost term in the effective action, was initially thought to be an independent symmetry. However, recent studies utilizing the BFV formalism have demonstrated that dual-BRST symmetry is not independent but can be derived via a canonical transformation in the ghost sector of BRST symmetry for U(1) gauge theory \cite{gaete,yale,rama}.

In this paper, we aim to make the BFV-BRST technique more accessible to non-experts by elucidating the connections between constrained systems, BRST, and dual-BRST symmetries. 

 
\section{BFV Generating Functional and Ordinary BRST Symmetry}\label{secbfv}
In this section, we have rendered the content of BFV-BRST technique more easily 
accessible to the non-experts by establishing the connection between the constrained systems, 
 BRST, dual-BRST symmetries and other forms of BRST symmetries for a prototypical first class system (generalized Rigid Rotor)\cite{thibe}.
  We consider  a prototypical first-class gauge-invariant dynamical system depending on the generalized canonical coordinates $(q^0,q^k,p_k)$, with $k=1,\dots,n$, defined by the Hamiltonian function
\begin{equation}\label{H}
H = U(q^k,p_k) + V(q^k)+q^0 T(q^k)\,,
\end{equation}
where $V(q^k)$ and $T(q^k)$ denote two given differentiable real functions.

The former represents an arbitrary physical potential, while the latter characterizes a first-class constraint condition imposed along the system dynamical evolution.
Still in equation (\ref{H}), we define further Latin indexes running though $i,j,k,l=1,\dots,n$, with repeated index summation convention always assumed. 
$U(q^k,p_k)\equiv\frac{R^{ijkl}T_iT_jp_kp_l}{2f^{ij}T_iT_j}
\,,$
with $f^{ij}=f^{ij}(q^k)$ denoting a symmetric nondegenerated two-form,
$R^{ijkl}$ given by
$R^{ijkl}(q^m)\equiv f^{ij}f^{kl} - f^{ik}f^{jl}$
and $T_i$ standing for the $T$ partial derivative with respect to $q^i$, i.e.,
$T_i\equiv\frac{\partial T}{\partial q^i},$
Using BFV approach, we introduce a pair of Grassmann anticommuting ghosts variables (${\cal{C}},\bar{\cal{C}}$) along with their respective canonically conjugated momenta ($\bar{\cal{P}},\cal{P}$)
with ghost numbers gh ${\cal{C}}=$ gh ${\cal{P}}=1=-$gh ${\bar{\cal{C}}}=-$gh $\bar{\cal{P}}$. 
The effective action in the extended 
phase space becomes
\begin{equation}
S_{eff}=\int dt\left ({\dot q}^i p_i + {\dot q}^0 p_0 + {\dot{\cal{C}}} \bar{\cal{P}} 
+{\dot{\bar{\cal{C}}}}{ \cal{P}} 
-H_\Psi \right )\,, \label{seff}
\end{equation}
 written in terms of
the extended Hamiltonian $H_\Psi$ obtained from (\ref{H}) by adding the extra ghost variables and gauge-fixing parts as
$H_\Psi =  U(q^k,p_k) + V(q^k) + \left\{\Omega_b,\Psi\right\}$,
with $\Omega$ denoting the BRST charge.
The symmetry generator for the system is constructed as
$\Omega_b=i\left({\cal{C}}T(q^k)+ {{\cal{P}}}p_0 \right )$
It generates the following ordinary BRST transformations
\begin{equation}\label{deltab}
\begin{array}{llll}
 \delta _b q^i =0\,,\quad \quad & \delta _b q^0 =-{\cal{P}}\,,\quad\quad & 
\delta _b {\cal{C}}=0\,,\quad\quad &
\delta _b
{\bar{\cal{C}}}=p_0 \,, \\
\delta _b p_i= {\cal{C}}T_i\,,\quad \quad & \delta _bp_0 =0\,,\quad \quad  & \delta _b {\bar{\cal{P}}}=T\,, \quad\quad
&  \delta _b {{\cal
{P}}}=0 \,.  
\end{array}
\end{equation}
Choosing the gauge fixing fermion as 
 \begin{equation}
\Psi= {\bar{\cal{P}}}q^0 +{\bar{\cal{C}}} \chi
\,,
\end{equation}
 we may write the quantum BRST invariant Hamiltonian as
\begin{equation}\label{hatH}
\hat{H} = U + V +  q^0 T +  p_0 \chi +i{\cal{C}}\left[T,\chi\right] {\bar{\cal
{C}}} +i{\cal P}\left[ p_0, \chi \right]
{\bar{\cal {C}}} + {\cal{P}}{\bar{\cal{P}}}
\,.
\end{equation}
 We consider  the standard gauge function as
 \begin{equation}\label{gf}
\chi=w^{-1}B+\frac{\xi}{2}p_0
\end{equation}
with the definitions
$B=f^{ij}T_ip_j\,~~~\mbox{ and }~~~
w=f^{ij}T_iT_j\,.$
The specific standard form (\ref{gf}) leads to the extended gauge-fixed action
\begin{eqnarray}\label{Sext}
S_{ext}=\int dt
\left (
{\dot q}^i p_i + {\dot{{\cal C}}}{{\bar{\cal{P}}}} + {\dot{\bar{\cal C}}}{{{\cal{P}}}} - U(q^k,p_k) - V(q^k) -q^0T(q^k)\right. \\ \nonumber
\left. - \frac{\xi}{2}p_0^2 +p_0{\left( {\dot{q}}^0-w^{-1}B \right)}
+ {\cal C}{\bar{\cal{C}}} - {\cal P}{\bar{\cal{P}}}
\right )\,, 
\end{eqnarray}

 The BRST transformations, generated by their respective charges, leave the extended action (\ref{Sext}) invariant:\\
Ordinary BRST
\begin{equation}  
\begin{array}{llll}
 \delta _b q^i =0\,,\quad  & \delta _b q^0 =-{\cal{P}}\,,\quad & 
\delta _b {\cal{C}}=0\,,\quad &
\delta _b
{\bar{\cal{C}}}=p_0 \,, \\
\delta _b p_i= {\cal{C}}T_i\,,\quad  & \delta _bp_0 =0\,,  & \delta _b {\bar{\cal{P}}}=T\,, \quad
&  \delta _b {{\cal
{P}}}=0 \,,
\end{array}
\end{equation}
The BRST Charge is 
\begin{equation}\label{Omegab}
\Omega_b=i\left[{\cal{C}}T(q^k)+ {{\cal{P}}}p_0 \right ]\,, ~~~~ \mbox{ gh }\Omega_b=+1\,.
\end{equation}

 Anti-BRST:
  
\begin{equation}\label{antideltab}
\begin{array}{llll}
 \bar{\delta} _b q^i =0\,,\quad  & \bar{\delta} _b q^0 =-{\bar{\cal{P}}}\,,\quad & 
\bar{\delta} _b {\cal{C}}=p_0\,,\quad &
\bar{\delta} _b
{\bar{\cal{C}}}=0 \,, \\
\bar{\delta} _b p_i= -{\bar{\cal{C}}}T_i\,,\quad & \bar{\delta} _bp_0 =0\,,\quad  & \bar{\delta} _b {\bar{\cal{P}}}=0\,, \quad
&  \bar{\delta} _b {{\cal
{P}}}=-T \,,
\end{array}
\end{equation}
The anti-BRST Charge is 
\begin{equation}\label{antiOmegab}
\bar{\Omega}_b=i\left[-{\bar{\cal{C}}}T(q^k)+ {\bar{\cal{P}}}p_0 \right ]\,, ~~~~ \mbox{ gh }\bar{\Omega}_b=-1\,.
\end{equation}

\section{Dual-BRST Symmetry: Hamiltonian Approach}\label{secdual}
 Dual-BRST: 
\begin{equation}\label{deltad}
\begin{array}{llll}
 {\bar{\delta}} _d q^i =0\,,\quad \quad & {\bar{\delta}} _d q^0 =-{\bar{\cal{C}}}\,,\quad\quad & 
{\bar{\delta}} _d {\cal{C}}=T\,,\quad\quad &
{\bar{\delta}} _d
{\bar{\cal{C}}}=0 \,, \\
{\bar{\delta}} _d p_i= {\bar{\cal{P}}}T_i\,,\quad \quad & {\bar{\delta}} _dp_0 =0\,,\quad \quad  & {\bar{\delta}} _b {\bar{\cal{P}}}=0\,, \quad\quad
&  {\bar{\delta}} _d {{
{\cal P}}}=p_0 \,,
\end{array}
\end{equation}
 Dual-BRST charge:
\begin{equation}\label{Omegad}
\bar{\Omega}_d=i\left[{\bar{\cal{P}}}T(q^k)+ {\bar{\cal{C}}}p_0 \right ]
\,, ~~~~ \mbox{ gh }\bar{\Omega}_d=-1\,.
\end{equation}
 Anti-dual BRST:  
\begin{equation}\label{antideltad}
\begin{array}{llll}
 {\delta} _d q^i =0\,,\quad & {\delta} _d q^0 =-{{\cal{C}}}\,,\quad & 
{\delta} _d {\cal{C}}=0\,,\quad &
{\delta} _d
{{\bar{\cal{C}}}}=-T \,, \\
{\delta} _d p_i= -{{\cal{P}}}T_i\,,\quad  & {\delta} _dp_0 =0\,,\quad  & {\delta} _d {{\bar{\cal{P}}}}=p_0\,, \quad
&  {\delta} _d {{\cal
{P}}}=0 \,,
\end{array}
\end{equation}
Anti-dual Charge
\begin{equation}\label{antiOmegad}
{\Omega}_d=i\left[-{{\cal{P}}}T(q^k)+ {{\cal{C}}}p_0 \right ]
\,, ~~~~ \mbox{ gh }{\Omega}_d=+1\,.
\end{equation}
All charges introduced above in equations (\ref{Omegab}), (\ref{antiOmegab}), (\ref{Omegad}) and (\ref{antiOmegad}) are off-shell nilpotent.

It is worth noticing that the two dual symmetries, ${\bar{\delta}}_d$ and ${\delta}_d$, can be obtained respectively from $\delta_b$ and $\bar{\delta_b}$ by the shifting
\begin{eqnarray}\label{can1}
a:~~~{\cal{C}} \longrightarrow {\bar{\cal{P}}} \,, ~~~ {\bar{\cal{C}}} \longrightarrow {{\cal{P}}} \,, ~~~ {{\cal{P}}} \longrightarrow {\bar{\cal{C}}}
\,, ~~~   {\bar{\cal{P}}}  \longrightarrow
{\cal{C}} \\ \nonumber
\,,~~~~\mbox{ (anti-)BRST }\longrightarrow\mbox{ (anti-)dual-BRST }\,,
\end{eqnarray}
Whereas the anti- symmetries, $\bar{\delta}_b$ and ${\delta}_d$, can be generated respectively from  $\delta_b$ and ${\bar{\delta}}_d$ by
\begin{eqnarray}\label{can2}
b:~~~{\cal{C}} \longrightarrow  - {\bar{\cal{C}}} \,, ~~~ {\bar{\cal{C}}} \longrightarrow  {\cal{C}}   \,, ~~~ {{\cal{P}}} \longrightarrow {\bar{\cal{P}}} 
\,, ~~~  {\bar{\cal{P}}} \longrightarrow
- {{\cal{P}}} \\ \nonumber 
\,,~~~~\mbox{ (dual-)BRST }\longrightarrow\mbox{ (dual-)anti-BRST }\,.
\end{eqnarray}
Both equations (\ref{can1}) and (\ref{can2}) above represent canonical transformations in the sense of preserving the fundamental relations 
\begin{equation}\label{QR}
\begin{array}{rllrll}
\left[ q^k, p_l \right]_- &=~ i\,\delta^k_{\,l} \,,&~~~~& \left[ \cal{C}, \bar{\cal{P}} \right]_+ &=~ -i \,,\\ 
\left[ q^0, p_0 \right]_- &=~ i \,,&~~~~&\left[ {\bar{\cal{C}}}, {{\cal{P}}} \right]_+ &=~ -i \,.
\end{array}
\end{equation} 

 All the four introduced BRST charges  are fermionic operators and fully off-shell nillpotent, i.e.,
${\cal Q}_b^2=\bar{{\cal Q}}_b^2=\bar{{\cal Q}}_d^2={\cal Q}_d^2=0\,$. 
The two (anti-)BRST operators commute among themselves
$\left[{\cal Q}_b,\bar{{\cal Q}}_b\right]= 0
\,,$
as well as the (anti-)dual-BRST ones
$\left[\bar{{\cal Q}}_d,  {\cal Q}_d \right] = 0
\,,$
while we have a non-null anticommutator between the (anti-)BRST and (anti-)dual-BRST given by
\begin{equation}\label{W}
\left[  {{\cal Q}}_b  ,  \bar{{\cal Q}}_d  \right]=
\left[  \bar{{\cal Q}}_b  ,  {{\cal Q}}_d  \right]=
i(T^2+p_0^2) \equiv 2 {\cal W}
\,.
\end{equation}

 The defined bosonic quantity ${\cal W}$ represents a Casimir operator for the superalgebra generated by the BRST charges, as it commutes with all of them, and gives rise to a new bosonic transformation defined for an arbitrary function $F$ as
\begin{equation}
s_W F \equiv [ F, {\cal W} ]
\,. 
\end{equation}
 For the fundamental variables of the theory, the non-null $s_W$ tranformations 
 \begin{equation}
s_W p_i = T T_i\,,~~~~~~
s_W q^0 = -p_0\,,
\end{equation}
and constitute a new symmetry leaving the action invariant. 
Similarly to the BRST charges, the Casimir operator $\cal W$ is a constant of motion being conserved along time evolution modulo equations of motion, albeit being a bosonic quantity.

\section{Conclusions: }\label{conc}
The quantization of the prototypical first-class system along the functional BFV procedure has allowed us to realize a considerable set of forms of BRST transformations constituting symmetries at quantum level. We have seen that the dual and anti BRST symmetries can be freely interchanged among a total of eight possibilities connected by canonical transformations in a Hamiltonian framework. The BRST charges exhibit the Hodge theory properties and it is possible to define a further Casimir operator leading to an extra bosonic symmetry and closing a Lie superalgebra among the conserved symmetry generators.
The simplicity and generality of the chosen first-class prototypical system permits those interpretations to be extended to other similar models in the literature shedding light on previous controversies regarding the physical interpretation of the dual BRST symmetries. 
 In a similar fashion to some of the mentioned field theory models, the prototypical first-class system can also exhibits new BRST symmetries involving the Lagrange multipliers, Nakanish-Lautrup variable and ghosts.

\end{document}